\documentclass[preview]{elsarticle}
\pdfoutput=1 
\usepackage{amssymb}
\usepackage{amsfonts}
\usepackage{amsmath}
\usepackage{color}
\usepackage{ulem}
\usepackage[portuguese, english]{babel}
\usepackage[utf8]{inputenc}
\usepackage[T1]{fontenc}

\biboptions{sort&compress}

\author{Artur Amorim\corref{cor1}}
\ead{artur.sousa@fc.up.pt} 
\author{Miguel S. Costa}
\ead{miguelc@fc.up.pt}
\author{Robert C. Quevedo}
\ead[url]{carcasses@gmail.com}

\affiliation[1]{organization={Centro de F\'{\i}sica do Porto e Departamento de F\'{\i}sica e Astronomia da Faculdade de Ci\^encias da Universidade do Porto},
addressline={Rua do Campo Alegre 687}, postcode={4169-007}, city={Porto}, country={Portugal}}

 \cortext[cor1]{Corresponding author}

\title{Deeply Virtual Compton Scattering in Improved Holographic QCD}

\begin{document}

\begin{abstract}
	We present our current progress in the holographic computation of the scattering amplitude for Deeply Virtual Compton Scattering (DVCS) processe, as a 
function of the Mandelstam invariant $t$. We show that  it is possible to describe simultaneously the differential cross-section and total cross-section of DVCS data with a single holographic model for the pomeron. Using data from H1-ZEUS we obtained a $\chi^2_{dof} \sim 1.5$ for the best fit to the data.
\end{abstract}

\maketitle
\flushbottom
\section{Introduction}

From the groundbreaking work of BPST \cite{Brower:2006ea} till today, many articles have shown how the theoretical ideas of $AdS/CFT$ can be used to model experimental data where pomeron exchange dominates~\cite{Hatta:2007he, Cornalba:2008sp,   Brower:2008ix, Levin:2009vj,  Hatta:2009ra, Kovchegov:2009yj, Avsar:2009xf, Cornalba:2009ax,  Cornalba:2010vk,   Kovchegov:2010uk, Levin:2010gc, Brower:2010wf, Ballon-Bayona:2017vlm, Costa:2012fw, Costa:2013uia, Brower:2012mk, Anderson:2014jia, Nally:2017nsp, Ballon-Bayona:2015wra, Amorim:2018yod,  Hatta:2008st,  Brower:2009bh, Domokos:2009hm, Domokos:2010ma, Stoffers:2012zw, Koile:2014vca,  Kovensky:2018xxa, Lee:2018zud, Kovensky:2018gif, Mamo:2019mka, FolcoCapossoli:2020pks}.
  These works are tailored for a specific process and/or kinematic regime. On the other hand, we have shown in~\cite{Amorim:2021ffr} that a holographic model with few free parameters that describes successfully several processes dominated by pomeron exchange is indeed possible. The processes analysed were sensitive to the forward scattering amplitude, i.e. for Mandelstam $t = 0$. So it remains an open question if it is possible to extend previous results to processes where amplitudes with non-zero values of $t$ are necessary. One such example is Deeply Virtual Compton Scattering (DVCS), which is the focus of this letter.

DVCS is an exclusive Compton scattering process where the incoming photon has high virtuality. It has been studied throughly by the H1 and ZEUS collaborations as well at JLAB. While DIS allows the determination of  nucleon parton distribuition functions (PDFs),  e.g. the proton PDFs, DVCS data is important for the study of the generalized parton distribution functions (GDPs) \cite{ji_deeply_1997,radyushkin_nonforward_1997}, which are related to the correlation between the transverse and longitudinal components of the momentum of quarks and gluons inside the nucleons.

Holographic techniques have been already employed to describe DVCS data. In \cite{Costa:2012fw} the conformal and hard wall pomeron models have been used, while in \cite{Stoffers:2012ai} a holographic description of dipole-dipole scattering has been studied. DIS is connected to the forward Compton scattering amplitude via the optical theorem. Hence, our analysis of DVCS in this work is close to the one presented in \cite{Ballon-Bayona:2017vlm}. We will show that in order to include the description of DVCS data one needs to include extra parameters that are related to the holographic wave function of the proton and to the coupling dependence with the spin of the exchanged Reggeons.

\section{Holographic computation of DVSC amplitude}

\begin{figure}
  \centering
  \includegraphics[scale=0.5]{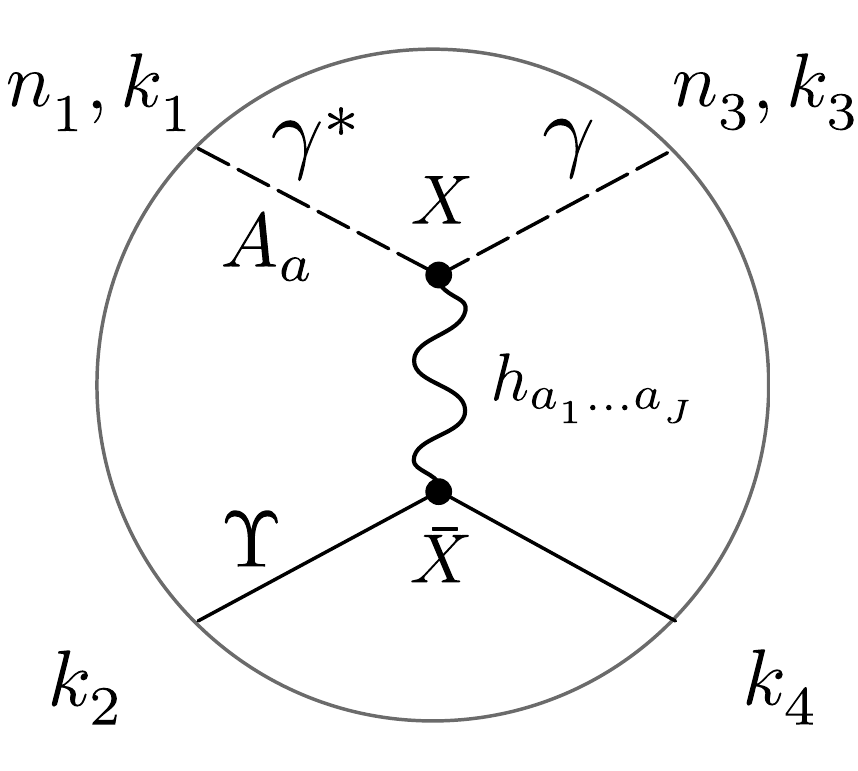}
  \caption[Tree level Witten diagram for the DVCS scattering amplitude]{Tree level Witten diagram associated with the computation of the amplitude $\mathcal{A}_J^{\lambda_1\lambda_3}$ of the DVCS process $\gamma^{*}p\to \gamma p$. The bottom lines represent the proton modeled by a scalar $\Upsilon$.}
  \label{fig:DVCS Witten diagram}
\end{figure}

We start by deriving holographic expressions for the differential cross-section $d \sigma / dt$ and total cross-section~$\sigma$ of the
$ \gamma^* p \to \gamma p $ DVCS process. In~\cite{Amorim:2021ffr} we computed the proton structure functions $F_2^p$ and $F_L^p$ and the total cross-section $\sigma\left(\gamma p \to X \right)$ by determining  the forward scattering amplitude of the process $\gamma^{*} p \to \gamma^{*} p$. Thus, the only difference between the DIS and the DVCS computation is that the outgoing photon is \textit{on-shell}. The associated Witten diagram is show in figure \ref{fig:DVCS Witten diagram}.
Hence we will consider the general case of $\gamma^* p \to \gamma^* p$ where the incoming photon has virtuality $Q_1^2$ while the outgoing photon has virtuality $Q_3^2$. At the end of the calculation we take $Q^3 \equiv k_3^2 \to 0$ and use the identity
\begin{equation}
\lim_{Q \rightarrow 0} f_Q (z) = 1\,, \quad \lim_{Q \rightarrow 0} \frac{\dot{f_Q}}{Q} = 0\,,
\label{eq:nonnormalizable_Q_0}
\end{equation}
where $f_Q$ is the non-normalizable mode of the bulk U(1) gauge field dual to the current $J^\mu = \bar{\psi} \gamma^\mu \psi$.
We take for the large $s$ kinematics of  $\gamma^*p\to \gamma^*p$ scattering the following momenta
\begin{align}
 \label{eq:NMC:kinematics}
&k_1=\left(\!\sqrt{s},-\frac{Q_1^2}{\sqrt{s}} ,0\right),
\qquad
 k_3=-\left(\!\sqrt{s},\frac{ q_\perp^2 -Q_3^2}{\sqrt{s}} , q_\perp \right),
\\
&k_2=\left(\frac{M^2}{\sqrt{s}},\sqrt{s} ,0\right),
\qquad
 k_4=-\left(\frac{M^2+ q_\perp^2}{\sqrt{s}},\sqrt{s} ,-q_\perp \right),
\nonumber
\end{align}
where $k_1$ and $k_3$ are,  respectively, the incoming  and outgoing photon momenta and $k_2$ and $k_4$ are, respectively, the incoming and outgoing proton momenta. 
The incoming and outgoing off-shell photons have the following polarization vectors
\begin{align}
    &n_1=
    \begin{cases}
      \left(0,0,\epsilon_\lambda \right), & \lambda=1, 2 \\
      \frac{1}{Q_1} \left( \sqrt{s}, \frac{Q_1^2}{\sqrt{s}}, 0, 0 \right), & \lambda = 3
    \end{cases}\,,\\
    &n_3=
    \begin{cases}
      \left(0,\frac{2 q_\perp \cdot \epsilon_\lambda}{\sqrt{s}},\epsilon_\lambda\right), & \!\lambda=1,2 \\
      \frac{1}{Q_3} \left(\sqrt{s}, \frac{Q_3^2+q_\perp^2}{\sqrt{s}}, q_\perp \right), & \!\lambda = 3
    \end{cases}\,,
    \label{eq:inPolarization}
\end{align}
where $\epsilon_1 =  (1,0)$ and $\epsilon_2 =  (0,1)$. 

To compute the Witten diagram in figure \ref{fig:DVCS Witten diagram}, consider  
the minimal coupling between the gauge field dual to the current $J^\mu = \bar{\psi} \gamma^\mu \psi$ 
and the higher spin field $h_{a_1 \dots a_J}$
\begin{align}
\label{eq:gauge_field_spin_j_coupling}
    k_J \int d^5 X \sqrt{-g} e^{-\Phi} F_{b_1 a} D_{b_2} \dots D_{b_{J-1}} F^{a}_{b_J} h^{b_1 \dots b_J},
\end{align}
where the field strength $F_{ab}$ is computed from the non-normalizable mode
\begin{equation}
A_a (X; k,\lambda)= n_a(\lambda)  f_Q(z) \, e^{ik\cdot x}\,.
\end{equation}
The coupling between the higher spin field and  the scalar field is
\begin{align}
\label{eq:scalar_field_spin_j_coupling}
  \bar{k}_J \int d^5  X \sqrt{- g} e^{- \Phi} \Upsilon D_{b_1} \dots  D_{b_J}  \Upsilon  h^{b_1 \dots b_J},
\end{align}
where 
\begin{equation}
 \Upsilon(X;k) =  \upsilon(z)\,e^{ik\cdot x}\,,
\end{equation}
is the normalizable mode describing  the proton state.
The computation of the scattering amplitude follows the same steps as the ones presented in~\cite{Amorim:2021ffr}, so we will not repeat it here. The scattering amplitude for $\gamma^* p \to \gamma p$ is
\begin{equation}
  \label{eq:amplitude gamma* gamma}
  \mathcal{A}^{\gamma^* p \to \gamma p}(s,t) = \sum_n  g_n(t)  s^{j_n(t)}\int dz \,e^{-(j_n(t)-\frac{3}{2})A} f_Q \,\psi_n(j_n(t))\,,
\end{equation}
where $A$ is the string frame warp factor and
\begin{equation}
 g_n (t)  = H(j_n(t)) \left[i + \cot \left(\frac{\pi j_n(t)}{2}\right)\right]  \frac{d j_n}{d t} \int dz \,e^{(-j_n(t)+\frac{7}{2})A} \upsilon^2 {\psi_n^{*} (j_n(t))} \, , \label{eq:dvcs_gn_def}
 \end{equation}
 with
 \begin{equation}
 H(J) = \frac{\pi}{2} \frac{k_{J} \bar{k}_{J}}{2^{J}} \, .
\end{equation}
The  amplitude (\ref{eq:amplitude gamma* gamma}) is the one where both photons have transverse polarizations, other combinations are subleading in $s$. To obtain this expression we  used   (\ref{eq:nonnormalizable_Q_0}).
The wavefunctions $\psi_n(j_n(t))$ are eigenfunctions of the effective Schr\"odinger potential~\cite{Ballon-Bayona:2017vlm}
\begin{align}
V_J(z) =  \frac{3}{2} \left( \ddot{A} - \frac{2}{3} \ddot{\Phi} \right) + & \frac{9}{4}  {\left( \dot{A} - \frac{2}{3} \dot{\Phi} \right)}^2 + e^{2A} (J-2) \left[ \frac{2}{l_s^2} \left(1+\frac{d}{\sqrt{\lambda}}\right) + \right. \notag \\
&\left.+ e^{-2A} \left( a \ddot{\Phi} + b\left( \ddot{A} - \dot{A}^2 \right) + c \dot{\Phi}^2 \right) +  \frac{J+2}{\lambda^{4/3}}  \right] ,
\end{align}
where $\lambda = e^\Phi$ is the 't Hooft coupling. The constants $l_s$, $a$, $b$, $c$ and $d$ are phenomenological parameters that will be fixed later by reproducing the best match with DVCS data. These parameters describe the analytic continuation of the equation of motion for the AdS spin $J$ field dual to twist 2 gluonic operators, in an expansion around $J=2$. This approximation gives the description of the graviton/pomeron Regge trajectory in the strong coupling approximation, in contrast with the weak coupling BFKL pomeron, whose expansion starts at $J=1$. The details of the derivation are given in \cite{Ballon-Bayona:2017vlm}.

To proceed we need to know the proton wavefunction in order to compute the $z$ integral in (\ref{eq:dvcs_gn_def}). In AdS/QCD, it is expected that baryons are dual to a configuration where three open strings are attached to a D-brane~\cite{witten_baryons_1998, Polchinski:2000uf}. So far it is not known how to derive the baryon spectrum from such a configuration. An acceptable holographic description of the proton would consist of a spectrum that matches the mass of the proton as well of the other hadrons in the same trajectory. Here we follow a phenomenological approach by approximating the combination $e^{3 A - \Phi}\upsilon^2$ by the delta function $\delta(z-z^*)$. The $z^*$ parameter is related, by dimensional analysis, to the inverse of the mass of the proton and  will be used  as a fitting parameter. By making this approximation we are assuming that  $e^{3 A - \Phi}\upsilon^2$  is null in the UV and the IR, and that it has a global maximum. This is expected if the spectrum of the baryons is associated with a Schrodinger problem whose ground state is the proton. This approach, as an example, has successfully described DVCS data in \cite{Costa:2012fw}. Another issue is the unknown functional form of the function $H(J)$ and its analytic continuation. In the next section we motivate a good ansatz for this expression, at least in the range $1 \lesssim J \lesssim 1.2$ that is relevant for the process here considered.

\section{AdS local coupling to the graviton trajectory}

\begin{figure}
  \centering
  \includegraphics[scale=0.5]{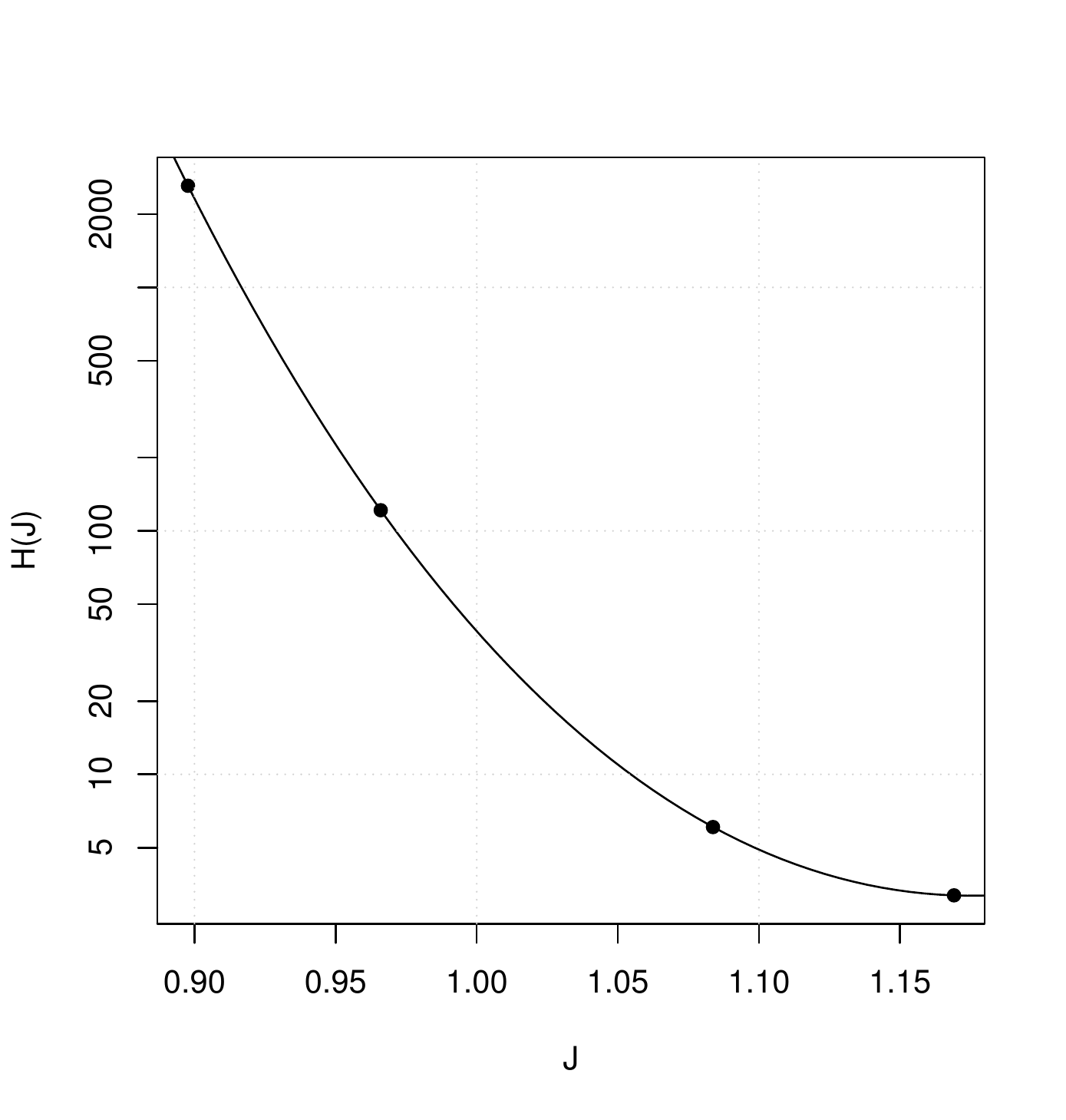}
  \caption[Reconstruction of the $H(J)$ function using the best parameters found in \cite{Ballon-Bayona:2017vlm} and $z^*=0.565$]{Reconstruction of the $H(J)$ function using the best parameters found in \cite{Ballon-Bayona:2017vlm} and $z^*=0.565$. The solid line represents the function $\exp \left(h_0 + h_1 (J-1) + h_2 (J-1)^2\right)$ with $h_0=3.70$, $h_1=-30.3$ and $h_2=89.1$.}
  \label{fig:dvcs:H(J)}
\end{figure}

To find a good ansatz for $H(J)$ let us first recall that in~\cite{Amorim:2021ffr}, based on the forward scattering amplitude for the process $\gamma^*p \to \gamma^* p$, the holographic formula for the proton structure function $F_2^p$ is given by
\begin{align}
&F_2^p(x, Q^2) = \sum_{n} \frac{ {\rm Im} \, g_n}{4 \pi^2 \alpha} \, x^{1-j_n}  Q^{2 j_n} \int dz \,e^{-\left(j_n-\frac{3}{2}\right)A}  \left( f_Q^2  +  \frac{\dot{f}_Q^{2}}{Q^2}      \right) \psi_n(z) \, ,
\end{align}
where $j_n$ and 
\begin{equation}
{\rm Im} \, g_n = H(j_n(0)) \frac{d j_n}{d t} \int d  z  \,e^{(-j_n(0)+\frac{7}{2}) A}  \upsilon^2 {\psi}_n^{*}  (j_n(0))
\end{equation}
are computed at $t=0$.
Using  (\ref{eq:dvcs_gn_def}) and the approximation $e^{3 A - \Phi}\upsilon^2 \sim \delta(z-z^*)$, we can compute $H(J)$ for any value of $j_n(t)$ with the result
\begin{equation}
  \label{eq:dvcs:H(J) vs g_n(t)}
  H(j_n (t)) = \frac{{\rm Im} \, g_n(t)}{e^{(-j_n (t) + \frac{1}{2})A(z^*)}\psi^*_n(j_n (t)) \frac{dj_n (t)}{dt}} \, .
\end{equation}
We can now plot $H(J)$ as a function of $J$ using
the $t=0$ values of ${\rm Im} \, g_n$, $j_n$, $\frac{dj_n}{dt}$ and $\psi_n$ for the different trajectories found in \cite{Ballon-Bayona:2017vlm} and  choosing $z^{*}$ to be around 1, since by dimensional analysis it should be of the order of the inverse mass of the proton (in GeV units). 
The black dots  shown in figure \ref{fig:dvcs:H(J)} represent the actual reconstructed values of $H(J)$ for each $n$.
In a logarithmic plot, $\log H (J)$ can be well approximated by a quadratic curve in $J$, i.e. $\log H(J) = h_0 + h_1 (J-1) + h_2 (J-1)^2$. We can then choose $z^{*}$ and the $h_i$ to get the best quadratic fit. For the best fit value of $z^* = 0.565$ we obtained the result presented  in figure \ref{fig:dvcs:H(J)}.

We can check if the resulting curve has physical meaning by considering the dependence  with $J$  of the
AdS local couplings  $\kappa(J)$ and $\bar \kappa (J)$, that define $H(J)$,
in the range $1 \lesssim J \lesssim 1.2$. According to the gauge/gravity duality such couplings are related to the OPE coefficient of a spin $J$ operator 
with two spin 1 operators (for the coupling with the EMG current operator) or spin 0 operators (for the coupling with a scalar, as we model the unpolarized proton). 
In particular, in the UV fixed point the OPE coefficient of two EMG current operators with the spin $J$ glueball operator associated with the pomeron 
trajectory vanishes, as only Wick contractions contribute.
Of course, QCD is not a free theory and perturbative corrections should be taken into account. Instead, to check if the proposed curve for $H(J)$ 
is reasonable let us consider the $O(N)$ vector model, which is also free in  the UV where the corresponding  OPE coefficient is non-vanishing \cite{Sleight:2017fpc}. In this model we can retrieve a similar shape of $H (J)$ as the one in figure \ref{fig:dvcs:H(J)} from the three-point bulk vertex coupling between massless higher spin fields of spin $s_1$, $s_2$  and $s_3$ in type A minimal higher-spin theory in $AdS_{d+1}$. The coupling, as a function of $s_1$, $s_2$ and $s_3$, is given by
\begin{equation}
g_{s_1,s_2,s_3} = \frac{\pi^{\frac{d-3}{4}} 2^{\frac{3d-1+s_1+s_2+s_3}{2}}}{\sqrt{N}\Gamma(d+s_1+s_2+s_3-3)} \sqrt{\frac{\Gamma(s_1 + \frac{d-1}{2})}{\Gamma(s_1 + 1)}\frac{\Gamma(s_2 + \frac{d-1}{2})}{\Gamma(s_2 + 1)}\frac{\Gamma(s_3 + \frac{d-1}{2})}{\Gamma(s_3 + 1)}}  \, .
\end{equation}
 After approximating the gamma functions in the resulting expression with the Stirling formula and expanding up to quadratic order around $s_3=J = 1$, we obtain a 
 function $H(J)$ consistent with our ansatz. Because we are comparing different theories caution should be taken. However, our point is that the overall function shapes does not change, and beyond the fixed point, corrections may be well captured in suitable redefinitions of the $h_i$ parameters in our ansatz. In the next section we test our hypothesis against DVCS experimental data.

\section{Data analysis and results}
Now that we have a reasonable parameterisation of $H(J)$ we proceed to find the best values for the pomeron kernel parameters $l_s$, $a$, $b$, $c$ and $d$,
 and the constants $h_0$, $h_1$ and $h_2$ in the $H(J)$ ansatz. The optimal set of parameters is found by minimising the $\chi^2$ statistic
\begin{equation}
  \label{eq:global chi2 definition}
  \chi_g^2=\chi_{\sigma}^2  +\chi_{\frac{d \sigma}{dt}}^2 \, ,
\end{equation}
which is the sum of the $\chi^2$ for $\sigma\left(\gamma^* p \to \gamma p\right)$ and the $\chi^2$ for $\frac{d\sigma\left( \gamma^* p \to \gamma p \right)}{dt}$. 
As usual, for a given observable $O \in \{\sigma\left(\gamma^* p \to \gamma p\right), \frac{d\sigma\left( \gamma^* p \to \gamma p \right)}{dt} \}$, the respective $\chi^2$ function is defined as
\begin{equation}
  \chi^2_{O} \equiv \sum_n \left(\frac{O_n^{\text{pred}} - O_n}{\delta O_n}\right)^2 \, ,
\end{equation}
with $O^{\text{pred}}$ being the predicted theoretical value and $\delta O$ the experimental uncertainty. The sum goes over the available experimental points.
The total and differential cross-section data used is the combined one from H1-ZEUS available in \cite{aaron_deeply_2009,chekanov_measurement_2009}.

The DVCS differential cross-section is given by
\begin{align}
    \frac{d\sigma}{dt} = \frac{1}{16 \pi s^2} \frac{1}{2} \sum_{\lambda_1,\lambda_3 = 1}^2 \left|A^{\lambda_1,\lambda_3}\left( s, t \right) \right|^2 =\frac{1}{16 \pi^2 s^2} \vert \mathcal{A}^{\gamma^*\gamma}(s,t) \vert^2,
\end{align}
where we average over the incoming photon polarization and the scattering amplitude is given by equation (\ref{eq:amplitude gamma* gamma}).
The total cross-section is just the integral of the above
\begin{equation}
\label{eq:dvcs_sigma}
  \sigma = \int_{-1}^0 dt \, \frac{d \sigma (t)}{dt} \,,
\end{equation}
where the integration range $-1 \leq t \leq 0$ comes from the data. 

To solve the $\chi^2$ minimisation problem we developed the $\tt{HQCDP}$ $\tt{R}$ package.
The Schrodinger problem associated to the gluon kernel is solved with $N=400$ Chebyshev points. To compute the differential and total cross-sections efficiently we divided the  interval $-1 \leq t \leq 0$ in 20 pieces of length 0.05 and computed the differential cross-section for each point. From these values we created a spline interpolation function that can be used to predict the differential cross-section values and the total cross-section through equation (\ref{eq:dvcs_sigma}). We also make use of the $\tt{REDIS}$ \textit{in memory} database to avoid redoing expensive computations. The code also makes use of multiple cores, if available, in order to compute in parallel the integrals that appear in (\ref{eq:amplitude gamma* gamma}) for different kinematical points. The present results were found using a node in a High Performance Computing (HPC) cluster with 16 cores.

\begin{figure}[t!]
  \centering
  \includegraphics[scale=0.5]{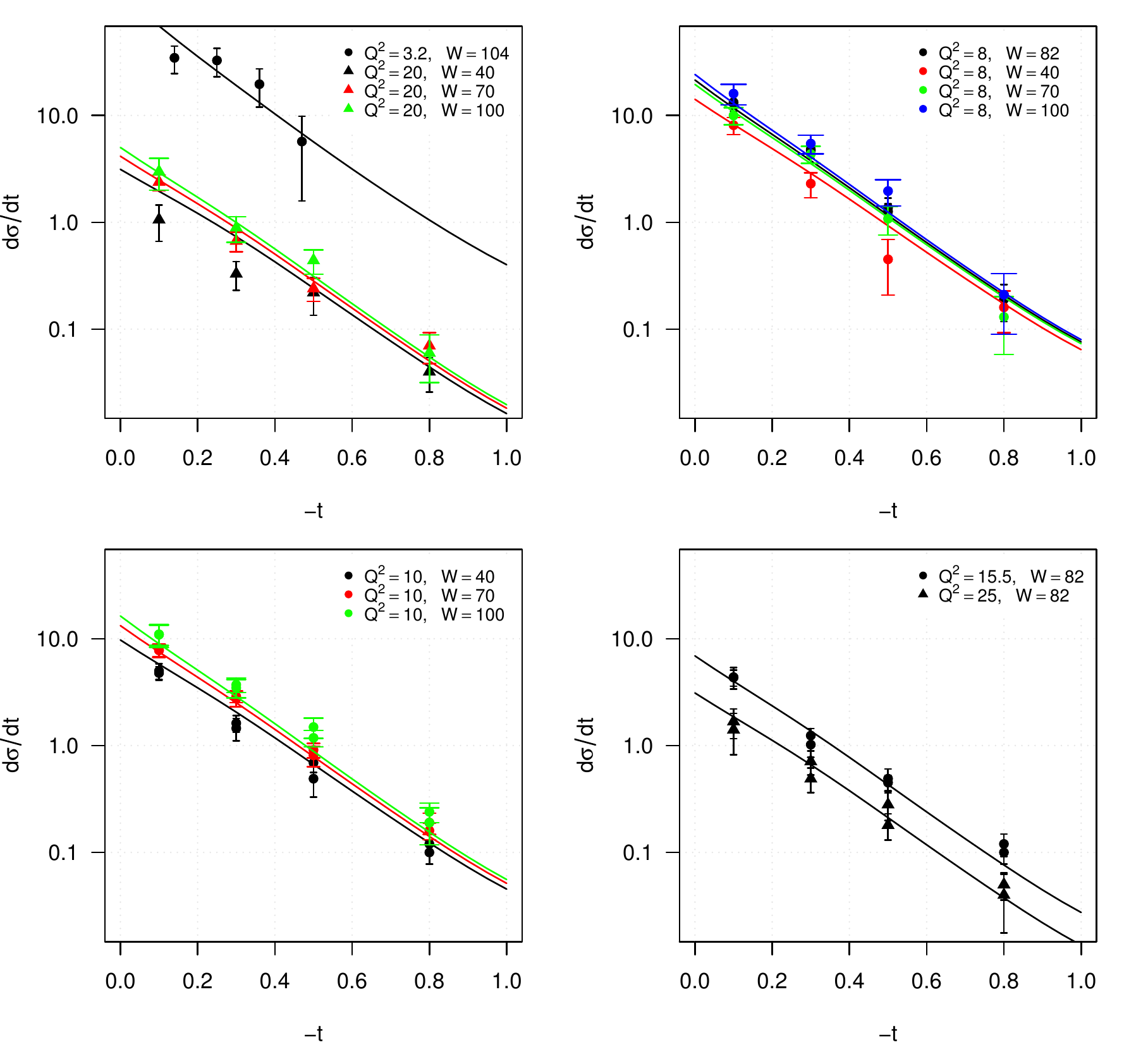}
  \caption[Predicted vs. experimental values of the differential cross-section $\frac{d \sigma(t)}{dt}$ for DVCS]{Predicted vs. experimental values of the differential cross-section $\frac{d \sigma(t)}{dt}$ for DVCS. Different gray levels correspond to different combinations of $Q^2$ and $W$ as described in the legends. Here $Q^2$ and $t$ are in $\text{GeV}^2$, $W$ in $\text{GeV}$ and $\frac{d \sigma}{d t}$ is in $\frac{\text{nb}}{\text{GeV}^2}$. }
  \label{fig:dvcs:DVCSDSigma}
\end{figure}

\begin{figure}[h!]
  \centering
  \includegraphics[scale=0.5]{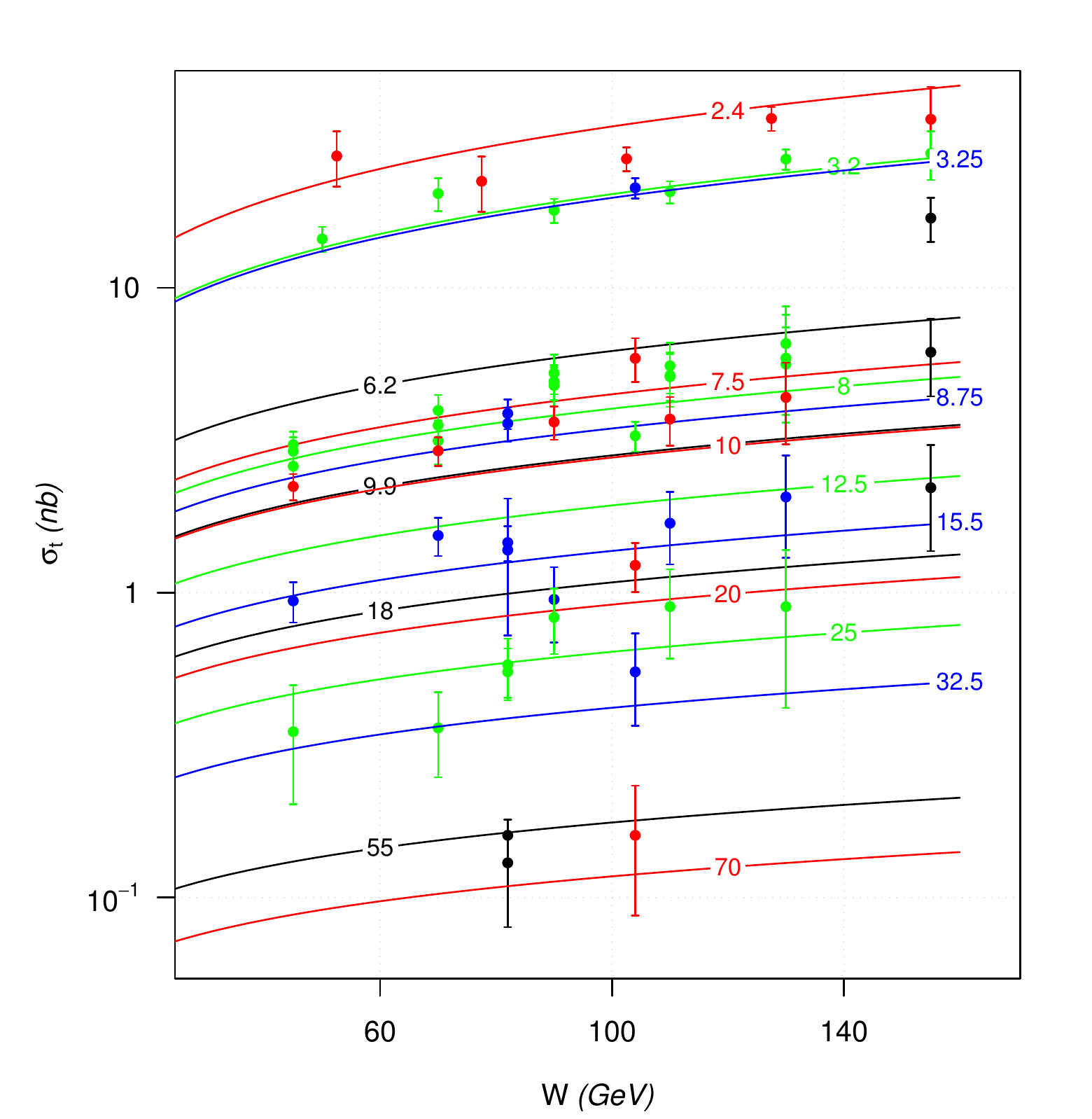}
  \caption[Predicted vs. experimental values of the total cross-section $\sigma$ of DVCS]{Predicted vs. experimental values of the total cross-section $\sigma$ of DVCS. Small numbers attached to lines and points of the same grey level indicate the respective value of $Q^2$ in $\text{GeV}$.}
  \label{fig:dvcs:DVCSSigma}
\end{figure}

\begin{figure}[h!]
  \centering
  \includegraphics[scale=0.5]{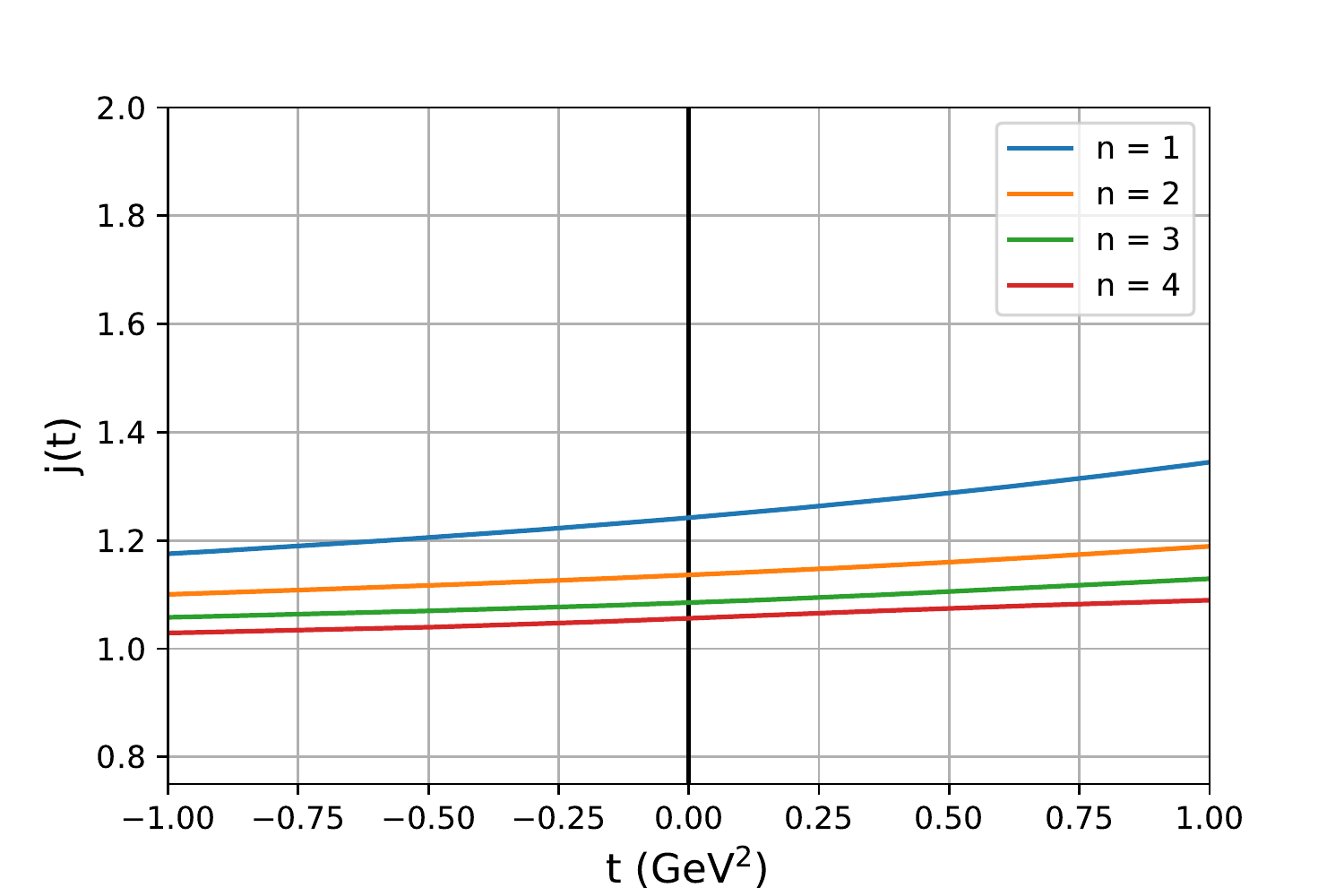}
  \caption{Regge trajectories obtained from our model and their intercepts. They were obtained using the kernel parameters of table~\ref{tab:DVCS DIS best values}. The plot includes the kinematical range
   $0 \lesssim-t \lesssim1\ ({\rm GeV}^2)$ as in the data used in this work.}
  \label{fig:regge_trajectories}
\end{figure}

The best fit to the data we have found has a $\chi^2_{dof} \sim 1.5$ with the parameter values of table~\ref{tab:DVCS DIS best values}. The individual values of $\chi^2_{dof}$ for the total and differential cross-section experimental data are $1.8$ and $1.3$ respectively, meaning that the model offers a good description of both processes. The comparison of the theoretical predictions against the experimental data can be seen in figures \ref{fig:dvcs:DVCSDSigma} and \ref{fig:dvcs:DVCSSigma}. Figure~\ref{fig:regge_trajectories} plots  the first four Regge trajectories obtained with the kernel parameters of table~\ref{tab:DVCS DIS best values}. For the data range $0 \lesssim-t \lesssim1\ ({\rm GeV}^2)$ the Reggeon spin 
is in the  range $1 \lesssim j_n(t) \lesssim 1.2$.
 
\begin{table}[h!]
  \centering
  \begin{tabular}{|c||c||c||c||c||c|}
    \hline 
    Kernel parameters & Extra parameters & Intercepts\\
    \hline 
    \hline 
    $a = -4.55$      & $h_0 = 4.74$  &  $j_0=1.24$ \\
    \hline 
    $ b = 0.980$     & $h_1 = -35.9$ &  $j_1=1.13$ \\
    \hline 
    $ c = 0.809$     & $h_2 = 142$  &  $j_2=1.08$\\
    \hline 
    $d = -0.160$     & $z^* = 0.296$ &  $j_3=1.05$\\
    \hline 
    $l_{s} = 0.153$ &--  &--  \\
    \hline 
  \end{tabular}
  \caption[Best fit parameters for the DVCS data fit]{The 9 parameters for our best fit and the intercept of the  first four pomeron trajectories. All parameters are dimensionless except $z^*$ and $l_s$ which are in $\rm{GeV}^{-1}$.}
  \label{tab:DVCS DIS best values}
\end{table}


To conclude, we have shown that the holographic model presented in~\cite{Amorim:2021ffr} can be extended to include a quantitative description of total and differential cross-sections of DVCS data from H1-ZEUS.  One should also include the kernel of the twist 2 fermion operators, as suggested by Donnachie and Landshoff~\cite{Donnachie:1998gm} and study the importance
of the exchange of meson trajectories  on the results above. A first step towards the inclusion of the twist 2 fermion operators has been done~\cite{Amorim:2021gat}.

\section*{Acknowledgments}

This research received funding from the Simons Foundation grant 488637  (Simons collaboration on the Non-perturbative bootstrap). 
Centro de F\'\i sica do Porto is partially funded by Funda\c c\~ao para a Ci\^encia e a Tecnologia (FCT) under the grant
UID-04650-FCUP.
 AA is funded by FCT under the IDPASC doctorate programme with the fellowship  PD/BD/114158/2016.

\bibliographystyle{elsarticle-num}
\bibliography{exp_data.bib, holography.bib, QCD.bib, glueballs.bib, books.bib, HVQCD.bib}
\end{document}